\documentstyle[aaspp4]{article}

\newcommand{\lc}{l_{\rm c}}
\newcommand{\vc}{V_{\rm c}}
\newcommand{\lgibs}{l_{\rm g}}
\newcommand{\st}{S_{\rm T}}
\newcommand{\Sl}{S_{\rm L}}

\begin{document}

\title{Can Deflagration-Detonation-Transitions occur in Type Ia Supernovae?}

\author{J. C. Niemeyer}
\affil{University of Chicago, Department of Astronomy and Astrophysics,\\
5640 S. Ellis Avenue, Chicago, IL 60637}

\begin{abstract}
The mechanism for deflagration-detonation-transition (DDT) by
turbulent preconditioning, suggested to explain the possible occurrence
of delayed detonations in Type Ia supernova explosions, is argued to be conceptually
inconsistent. It relies crucially on diffusive heat losses of the
burned material on macroscopic scales. Regardless of the
amplitude of turbulent velocity fluctuations, the typical gradient
scale for temperature fluctuations is shown to be the laminar flame width or
smaller, rather than the factor of thousand more required for a
DDT. Furthermore, thermonuclear flames cannot be fully quenched in regions much
larger than the laminar flame width as a consequence of their simple
``chemistry''. Possible alternative explosion scenarios are briefly discussed. 
\end{abstract}

\keywords{hydrodynamics, stars: supernovae: general}

\section{Introduction}
The delayed detonation scenario for Type Ia supernova (SN Ia)
explosions asserts that a slowly accreting Chandrasekhar mass C+O
white dwarf undergoes 
a thermonuclear explosion in two distinct modes: an initial turbulent
deflagration (flame) phase that preexpands the star, allowing the
abundant production of intermediate mass isotopes observed in SN Ia
spectra, followed by a detonation that accounts for the 
high material velocities and the strength of the explosions (Khokhlov
1991; Woosley \& Weaver 1994). Both modes are assumed to be linked by
a deflagration-detonation-transition (DDT) that occurs 
either during the first expansion phase or after a partial recollapse
of the star (Arnett \& Livne 1994). The background density of the DDT,
$\rho_{\rm t}$, is often referred to (H\"oflich et al. 1996; Nomoto et al. 1997)
as the leading candidate for the physical parameter corresponding to
the observed correlation of peak luminosity and light curve shape
(Pskovskii 1977; Phillips 1993). 

Despite the apparent success of one-dimensional delayed detonation
models in reproducing many features of observed SN Ia spectra and
light curves (H\"oflich et al. 1996), a quantitative investigation of
DDTs in supernovae has begun only recently. Both dimensional analysis
and numerical simulations   indicate that a turbulent thermonuclear
flame front driven on large scales by the Rayleigh-Taylor (RT)
instability falls short of sonic propagation by at least one order of
magnitude (Niemeyer \& Hillebrandt 1995; Khokhlov 1995; Reinecke et
al. 1998a), making early proposals for DDT by direct shock formation
seem implausible. An alternative route to detonations in supernovae
involving local flame quenching and microscopic turbulent mixing was
recently proposed (Khokhlov et al. 1997b; Niemeyer \& Woosley
1997). It is based on the induction time gradient mechanism (Zeldovich
et al.~1970), first applied to detonations in supernovae by Blinnikov
\& Khokhlov (1986, 1987). This mechanism requires a sufficiently large
region of unburned fluid to be preconditioned in a manner that
establishes a uniform temperature gradient across it. For
predominantly temperature dependent reaction rates, the temperature
gradient can be mapped onto a gradient of induction times and hence
onto a phase velocity for the spontaneous burning wave that sweeps
across the region. When certain criteria regarding the fuel mixture
fraction and the size of the region are met
(Sec.~\ref{prerequisites}), the pressure released by the burning wave
may form a  self-sustaining reaction-shock complex, i.e. a
detonation. Note that no microscopic transport nor high fluid
velocities are needed for the runaway to detonation. Instead, the
problems of those proposals for DDT that need supersonic turbulent
flame speeds are now entirely passed on to the preconditioning of the
gradient regiion. It is well known that the gradient mechanism is very
sensitive to even small temperature non-uniformities (Blinnikov \&
Khokhlov 1986, 1987; Woosley 1990; Niemeyer \& Woosley 1997). Owing to
the absence of walls or obstacles, the 
preferred locations for DDT in confined systems, preconditioning in
supernova explosions can only be attributed to mixing in an unconfined
turbulent flow field. It will be shown in Sec.~(\ref{noquench}) that
under these conditions, successful preconditioning entails a degree of
synchronicity that is irreconcilable with subsonic turbulence.

In this paper, it will be argued that turbulent mixing in large
systems does not, in general, give rise to uniform gradients on large
scales (Sec.~(\ref{noquench})). Furthermore, even if locally isolated
regions are considered, the robustness of thermonuclear flames with
respect to turbulent quenching disfavors the emergence of sufficiently
well-mixed regions for DDT. We conclude
that unless we are missing an important piece of information, the  
physics of unconfined  turbulent thermonuclear flames appears to allow
transitions to a detonation only in the case of rare fluctuations
instead of providing a robust framework for DDT. Some alternative
explosion scenarios will be outlined in Sec.~(\ref{scenarios}). 

\section{Prerequisites for deflagration-detonation-transitions in
supernovae}
\label{prerequisites}

Assuming that there are no natural sources of shocks in the turbulent
flame brush of a supernova explosion (for a possible exception, see
the description of active turbulent combustion (ATC) in
Sec.~(\ref{scenarios})), such as corners or obstacles in 
terrestrial combustion experiments, the only  way to create
a pressure spike that turns into a detonation is by burning a certain
critical volume, $\vc \sim \lc^3$,  of fuel within a time
comparable to or less than its sound-crossing time, $t_{\rm s}(\lc) \sim
\lc/u_{\rm s}$. This can be achieved in two very different
ways. On the one hand, turbulent deformation of the flame surface can,
in principle, create a sufficiently large flame surface area to burn a
given volume in an arbitrarily short time. This idea is the motivation
behind the fractal model (Woosley 1990). However, a simple argument shows 
that this can only occur in rare fluctuations as long as the
steady-state turbulent flame velocity, $\st$, is subsonic, since the
statement of burning $\vc$ within $t_{\rm s}$ is equivalent to $\st
\sim u_{\rm s}$ if $\st$ is evaluated on the scale $\lc$. Given that
in the flamelet regime, the turbulent flame speed scales with the
turbulent velocity fluctuations on each scale, $\st(l) \sim v(l)$, and
that the latter is bound from above by the (subsonic) terminal rise
velocity of buoyant RT bubbles, it is clear that this mechanism is an
unlikely candidate for a robust DDT scenario (Niemeyer \& Woosley 1997).

On the other hand, detonations might be created via the well-studied
induction time gradient mechanism (Zeldovich et al. 1970; Lee et
al. 1978), whereby a combustion  
wave moving along a preconditioned temperature gradient coherently 
builds up a pressure wave that -- for sufficiently large
preconditioned volumes -- eventually turns into a detonation (for a
recent discussion, see Khokhlov et al. 1997a,b). The minimum size $\lc$
of the preconditioned region that gives rise to a detonation in white
dwarf matter was
derived numerically by Niemeyer \& Woosley (1997) and Khokhlov et al. (1997b).
It sensitively depends on the composition and density of the fluid;
however, for the purpose of this paper it is sufficient to note that in all 
cases, $\lc$ is larger than the laminar flame width $\delta$ by more
than three orders of magnitude. 

So far, the problem has merely been shifted from the fine-tuning of
the flame surface area within the critical region to that required to
precondition the temperature field. In both cases, the only tool
naturally available is subsonic buoyancy-driven turbulence. However,
as pointed out by Khokhlov et al. (1997a,b) and Niemeyer \&
Woosley (1997), it is possible in principle that for a given laminar flame speed
and width there exists a critical turbulence intensity such that
turbulent mixing can locally extinguish, or quench, the nuclear
reactions within the flame. If this were the case, turbulence might be
able to mix burned and unburned material and establish an
appropriately smooth temperature field. The details of this mixing
process, however, were not investigated in previous studies. 

Niemeyer \& Woosley (1997) and Khokhlov et al. (1997a,b) used the Gibson
length $\lgibs$, defined as the scale where the turbulent eddy velocity is
equal to the laminar flame speed, $v(\lgibs) \sim \Sl$, to postulate the
necessary conditions for flame quenching: if $\lgibs \le \delta$, the burning
regime changes from ``flamelet'' to ``distributed'' burning and
turbulence begins to appreciably affect the diffusion-reaction
structure of the flame. Only in the distributed burning regime can local flame
quenching take place. The criterion above was later shown to be
equivalent to a definition of the flamelet regime based on the
relative strengths of turbulent and thermal diffusivities (Niemeyer \&
Kerstein 1997). Note that for Prandtl numbers $Pr=\nu/\kappa$,
defined as the ratio of viscosity and thermal conductivity, below
unity this criterion is in conflict with conventional flamelet theory 
(Peters 1984) which relies on a comparison of the Kolmogorov length
and $\delta$; according to this definition, thermonuclear flames with
$Pr \ll 1$ would
never be anywhere near the flamelet regime. Recent numerical
experiments favor the modified flamelet definition as opposed to the
conventional one (Niemeyer et al. 1999).

Intriguingly, the transition from flamelet to distributed burning, and
hence the first chance for turbulence to create large islands of
preconditioned material, approximately takes place at the right
transition density for DDT, $\rho_{\rm t} \sim 10^7$ g cm$^{-3}$,
inferred from one-dimensional explosion models (Niemeyer \& Woosley
1997). It could therefore 
provide the switch that triggers detonations in the late phase of
supernova explosions, replacing a free model parameter with a
physical one. To conclude this section, the combination of the
gradient mechanism for DDT and the transition from flamelet to
distributed burning at a density of $\sim 10^7$ g cm$^{-3}$ may
explain the bulk of SN Ia observations, but it crucially hinges on the
existence of a mechanism for turbulent preconditioning of a region
much larger than the laminar flame width at that density. 

\section{Failure of turbulent flame quenching and macroscopic
preconditioning} 
\label{noquench}

Consider first an infinite fluid dynamical system, containing a
passive scalar field that changes from zero to a finite value across the
domain of interest, and is subject to self-similar turbulent mixing in
the center of the domain. Assuming for now that
turbulence or expansion manage to fully extinguish nuclear burning,
this is a reasonable description of the temperature field $T$ in the
turbulent flame brush, since
the length scales we are interested in, $l \sim \lc$, are much smaller
than the stellar radius and the turbulence on these scales had ample
time to establish a self-similar cascade. Under these conditions, the
temperature fluctuation amplitudes obey 
Kolmogorov scaling, $T(l) \sim l^{1/3}$. The temperature field becomes
a smooth function at the heat diffusion scale given by $l_{\rm d} \sim L Re^{-3/4}
Pr^{-3/4}$ for $Pr < 1$, where $Re$ is the Reynolds number and $L$ is
the integral scale of the turbulence. Under conditions typical for the onset
of distributed burning in SN Ia models, $L \approx 
10^7$ cm, $Re \approx 10^{14}$, and $Pr \approx 10^{-4}$, the heat
diffusion scale is $l_{\rm d} \sim 10^{-1/2} $cm $\sim \delta \ll l_{\rm c}$.
Evidently, increasing the turbulence intensity (and thus $Re$) decreases 
the largest length scale where $T$ can be considered smooth, rather
than increasing it. Regardless of the amplitude of 
large-scale turbulent velocity fluctuations, turbulent mixing is
inherently unable to provide uniformly mixed regions on
macroscopic scales $\lc$. Dropping the simplification of treating $T$
as a passive scalar further strengthens this statement, as burning
strongly enhances temperature fluctuations on scales $\sim \delta \ll
l_{\rm c}$. 

The question of DDT in the presence of temperature fluctuations was
recently investigated numerically by Montgomery et al.~(1998). It was
found that perturbation amplitudes of 10-15\% are sufficient to
divide the gradient region into subregions, each of which would need
to have the size of the unperturbed critical length $\lc$ in order to
give rise to a detonation. However, this study optimistically
assumed that a constant temperature gradient of order $\lc^{-1}$
exists initially and is subsequently perturbed by turbulent
fluctuations on smaller scales. As argued above, these initial
conditions are inconsistent with a self-similar turbulent mixing
region. 

One may also drop the assumption of self-similarity by looking at the
special case of a locally isolated fluid element, recognizing that while
these are not typical regions of a turbulent flow, a small number of them may
be realized on statistical grounds. Consider, for
instance, a single large eddy of size $\sim \lc$ with little or no
entrainment of material from the outside. In this case, the passive
scalar is mixed microscopically over the entire region after
approximately one eddy turn-over time $\tau_{\rm eddy}(\lc)$. This
situation would, in fact, 
give rise to suitable preconditioning for DDT if burning could be
inhibited during the mixing process; otherwise, small scale
fluctuations on the scale $\delta$ are continually resupplied. The
remaining question is thus: can turbulence quench nuclear reactions in
a region as large as $\lc \gg \delta$? More specifically, can the
burning products contained in $\vc$ be cooled sufficiently such that
the burning time scale $\tau_{\rm b} \sim \dot w^{-1}$, where $\dot w$
is the fuel consumption rate, is larger than $\tau_{\rm eddy}(\lc)$ everywhere 
within $\vc$?

The answer is no, provided that heat loss to the environment is
negligible and the flow is subsonic. For simplification, we shall
concentrate on carbon burning alone, since it represents the fastest
reaction and its extinction is a
necessary (and sufficient) condition for flame quenching. Ignoring the
small density change across the flame, the carbon burning rate $\dot
w_{\rm C+C}$ depends only on temperature and carbon mixture fraction. Note further
that because of electron degeneracy, heat diffuses many orders of
magnitude more rapidly than nuclei, so that we can safely assume
that carbon is non-diffusive. Consequently, flame quenching can only
occur by diffusive cooling ($p$d$V$-cooling is irrelevant because the
flow is to a very good approximation incompressible). Turbulence
affects the efficiency of diffusion by straining the flame and thus
steepening the temperature gradients. For temperature gradients of order
$\delta^{-1}$, the diffusion time scale, $\tau_{\rm d}(\delta) \sim
\delta^2/\kappa$, is by definition of $\delta$ comparable to the burning
time scale $\tau_{\rm b} \sim \dot w_{\rm C+C}^{-1}$. For gradients
larger than $\delta^{-1}$, diffusion is faster than burning throughout
most of the flame. However, it
can lead to full extinction only if the entire region of burning
products that it is connected with is also smaller than $\sim \delta$,
in which case heat can leak out to all sides and the products can be
cooled sufficiently to satisfy $\tau_{\rm b} \ll \tau_{\rm
d}$. Otherwise, if the flame is connected to a heat bath of burning
products larger than $\delta$, the temperature at the interface of
fuel and ash always remains fixed at the final product temperature,
keeping $\tau_{\rm b}$ small in its immediate vicinity, regardless of
the strain rate experienced by the flame. The total burning rate may
drop with respect to the flamelet regime, but fast nuclear burning is
never fully extinguished within the whole volume. 

According to these arguments, the only conceivable way to quench the flame in a
large volume $\vc$ is to stretch it into a thin filament with thickness
$\le \delta$ and curl it up such that it fills $\vc$. In order to prevent
unquenched burning in any part of $\vc$ before the onset of the
spontaneous runaway, this curling has to be completed in a time $t \ll
t_{\rm s}(\lc)$. Interestingly,
we now face the same problem as the fractal model described in the
previous section: the eddy velocity has to be supersonic in order to prepare the
runaway region before it is burned. Again, we are limited by the fact
that a subsonic process cannot set up conditions that are later
supposed to burn with a supersonic phase velocity.

The line of arguments above is supported by numerical (Poinsot et
al. 1991) and experimental (Shy et al. 1996) 
evidence that premixed chemical flames can only be quenched in the
presence of heat losses or complicated thermochemical effects, both of
which are absent in thermonuclear combustion. Further confirmation was
obtained with a one-dimensional calculation of a thermonuclear flame
subject to discrete multiscale remappings representing turbulent eddies
(Lisewski et al. 1999). The interaction of simple diffusion-reaction
flames with turbulence on the scale of the flame width was studied by
Niemeyer et al. (1999), demonstrating that local flame propagation is
nearly unaffected by turbulence even if the turbulence intensity is
comparable to the laminar flame speed.

\section{Alternative scenarios}
\label{scenarios}

If the initial deflagration phase fails to release enough energy to
unbind the star and no DDT takes place during the expansion, the star
pulses and eventually recontracts, revitalizing the turbulence by
compression (Arnett \& Livne 1994; Khokhlov 1995). During the 
pulse, the cut-off scale for temperature fluctuations $l_{\rm d}$ can
grow extremely large because turbulence is essentially frozen
in. At very low densities the flame width $\delta$ is
macroscopically large, allowing the formation of fluid regions which
-- if they survive the recontraction phase without disruption -- may
be suitably preconditioned for DDT later on. However, turbulent
entrainment of hot and cold material during the collapse will again
raise the amplitude
and lower the cut-off scale of temperature fluctuations. It
is impossible to say a priori whether the fluid is more likely to reignite
in the deflagration or detonation mode. While the extensive mixing
period during the pulse probably helps to create favorable conditions
for DDT, its benefits may well be erased by the enhanced turbulence
intensity during the recontraction. Moreover, the extremely
fine-tuned time synchronization required for the gradient mechanism
for DDT seems to be as unnatural in the pulsational mode as in the direct
one. 

An additional problem of the pulsational delayed detonation scenario
was pointed out by Niemeyer \& Woosley (1997): if a large pulse is
needed to achieve the required degree of homogeneity, what are the
observational counterparts of those events that barely unbind the star
but do not detonate? One may evade this problem by assuming that
turbulent deflagrations reliably fall short of releasing the binding energy 
of the white dwarf. This, however, is in conflict with the latest 
two-dimensional simulations that indicate a clear trend toward higher energy
release with increased numerical resolution (Hillebrandt et al. 1999).
These simulations employ a flame capturing algorithm
based on the level set method (Reinecke et al. 1998b) that shows the
emergence of more and more flame 
structure as the grid resolution is improved. For certain initial
conditions, the star clearly becomes unbound, yet no convergence of
total energy generation with respect to resolution has been achieved so
far. Should this trend continue, and ultimately be confirmed in
three-dimensional calculations, there is a realistic possibility that
turbulent deflagrations alone are sufficient to power the explosions
without the need for detonations. 

The simulations by Niemeyer et al. (1996) and Reinecke et al. (1998a)
further demonstrate that the role of the initial conditions for flame
ignition has not yet been fully explored. If the explosion is sparked
off at many disconnected points, the complexity of the flame surface
later on may easily exceed the surface area derived from the
non-linear growth of an initially smooth, RT unstable
interface (``dandelion model'', Niemeyer \& Woosley
1997). One-dimensional SN Ia models are unable to adequately
represent such effects.  

Finally, we can consider alternative routes to detonations that
do not rely on large scale preconditioning. One such possibility is
active turbulent combustion (ATC) (Kerstein 1996; Niemeyer \& Woosley 1997),
a runaway process of turbulent combustion that may occur as a
consequence of flame-generated turbulence on multiple scales. Scaling
arguments show that in the absence of an effective mechanism 
for stabilization, a runaway must ensue in any unconfined turbulent
flame brush (Kerstein 1996). It is possible that the non-linear stabilization
mechanism of the Landau-Darrieus (LD) instability by cusp formation
(Zeldovich 1966) is unstable with respect to finite amplitude
perturbations exerted by turbulent fluctuations, giving rise to an
increasingly more violent acceleration of the flame front that ends
only when compressibility effects become important. Cusp stabilization
of the LD instability may also break down at a critical expansion
ratio of burned and unburned material, as suggested by Blinnikov \&
Sasorov (1996). Practically, the
consequences of ATC would involve either nearly sonic turbulent combustion
or direct DDT by shock formation ahead of the combustion front. While undoubtedly
speculative, ATC is a promising mechanism for powerful SN Ia explosions
without the need for fine-tuning. Numerical experiments
designed to measure the relevance of ATC for thermonuclear flames are
underway.

\section{Conclusions}

This paper argues that the gradient mechanism for
deflagration-detonation-transitions (DDT), previously believed to be
the most realistic candidate to explain delayed detonations in Type
Ia supernovae (SN Ia), is inconsistent with the phenomenology of turbulent
mixing and combustion. Combining the inability of turbulence to
provide microscopic mixing over macroscopic length scales with the
robustness of thermonuclear flames with respect to quenching, the
establishment of sufficiently large regions with a nearly constant
temperature gradient is shown to be very unlikely. Both of these effects
can (and must) be verified by means of direct numerical simulations on
small scales. Work in this direction is in progress; first results of
flame-turbulence interactions on small scales can be found in
(Niemeyer et al. 1999). The argument above
holds as well for pulsational explosions, although here the
long intermediate period of diffusion dominated mixing may slightly
facilitate the preconditioning needed for DDT. 

Why, then, do one-dimensional explosion models with a slow deflagration
phase followed by a delayed detonation so successfully fit the
majority of SN Ia observations? Either we are missing an important
effect that robustly leads to a DDT or at least to a very fast
turbulent flame late during the explosion -- a noteworthy, albeit
speculative, possibility is active turbulent combustion (ATC) --
or 1D models get the right answer for the wrong reasons, because they
cannot accurately represent important multidimensional effects. An
example for the latter is the impact of multipoint ignition on the
development of the flame surface complexity, an effect that may well
lead to a strongly enhanced burning rate in the deflagration mode as
compared with the standard scenario. In any case, the success of both
the direct and the  
pulsational modes for DDT hinges on a deflagration phase that is much
slower than indicated by recent results of two-dimensional simulations. 

On the other hand, there is a trend toward higher energy release by the
turbulent flame if the numerical resolution is increased. So far, no
convergence of the total energy generation has been attained. If this
trend continues and is confirmed by three-dimensional calculations
with realistic subgrid-scale modeling, the possibility that the bulk
of Type Ia supernovae explodes without ever detonating must be taken
more seriously. 

To summarize, our analysis suggests that detonations may never take
place in SN Ia explosions. If they do, they probably need to be preceded by a
nearly sonic turbulent deflagration, in which case it may not be
possible to clearly distinguish deflagrations from detonations
observationally. ATC, multipoint ignition, higher than anticipated
energy release in the turbulent flame brush, or any combination
thereof may provide the required energy output to power the
explosion. 

\acknowledgements
This work is the result of many interesting discussions with Kendal
Bushe, Alan Kerstein, Wolfgang Hillebrandt, Bob Rosner, and Stan 
Woosley. I also wish to acknowledge helpful information from Sergej
Blinnikov, Greg Ruetsch, Martin Reinecke, and Martin Lisewski. This
research was supported in part by the ASCI Center on Astrophysical
Thermonuclear Flashes at the University of Chicago under DOE contract
B341495.


\begin{thebibliography}{99}

\bibitem{}
Arnett, D. and Livne E. 1994, \apj, 427, 330

\bibitem{}
Blinnikov, S. I. and Khokhlov A. M. 1986, Soviet Astron.~Lett., 12,
131

\bibitem{}
Blinnikov, S. I. and Khokhlov A. M. 1987, Soviet Astron.~Lett., 13, 364

\bibitem{}
Blinnikov, S. I. and Sasorov, P. V. 1996, Phys. Rev. E, 53, 1

\bibitem{}
Hillebrandt, W., Reinecke, M., and Niemeyer, J. C. 1999,
Comp. Phys. Comm., in press 

\bibitem{}
H\"oflich, P., Khokhlov, A. M., Wheeler, J. C,. Phillips, M. M.,
Suntzeff, N. B., and Hamuy M. 1996, \apjl, 472, L81

\bibitem{}
Kerstein, A. R. 1996, Comb. Sci. Tech., 118, 189

\bibitem{}
Khokhlov, A. M. 1991, \aap, 246, 383

\bibitem{}
Khokhlov, A. M. 1995, \apj, 449, 695

\bibitem{}
Khokhlov, A. M., Oran E. S., and Wheeler, J. C. 1997a, Combust. Flame,
108, 503

\bibitem{}
Khokhlov, A. M., Oran E. S., and Wheeler, J. C. 1997b, \apj, 478, 678

\bibitem{}
Lee, J. H. S., Knystautas, R., and Yoshikawa, N. 1978, Acta
Astronaut., 5, 971

\bibitem{}
Lisewski, A. M., Hillebrandt, W., Woosley, S. E., Niemeyer, J. C., and
Kerstein, A. R. 1999, in preparation

\bibitem{}
Montgomery, C. J., Khokhlov, A. M., and Oran, E. S. 1998,
Combust. Flame, 115, 38

\bibitem{}
Niemeyer, J. C. and Hillebrandt, W. 1995, \apj, 452, 769

\bibitem{}
Niemeyer, J. C., Hillebrandt, W., and Woosley, S. E. 1996, \apj, 471, 903

\bibitem{}
Niemeyer, J. C. and Kerstein, A. R. 1997, New Astronomy, 2, 239

\bibitem{}
Niemeyer, J. C. and Woosley, S. E. 1997, \apj, 475, 740

\bibitem{}
Niemeyer, J. C., Bushe, W. K., and Ruetsch, G. R. 1999, \apj, in press

\bibitem{}
Nomoto, K., Iwamoto, K., and Kishimoto, N. 1997, Science, 276, 1378

\bibitem{}
Peters, N. 1984, Prog. Energy Combust. Sci., 10, 319

\bibitem{}
Phillips, M. M. 1993, \apj, 413, L105.

\bibitem{}
Poinsot, T., Veynante, D., and Candel, S. 1991, J. Fluid Mech., 228,
561

\bibitem[Pskovskii 1977]{1977SvA....21..675P} 
Pskovskii, I. P. 1977, Soviet Astronomy, 21, 675 

\bibitem{}
Reinecke, M., Hillebrandt, W., and Niemeyer, J. C. 1998a, \aap, in
press

\bibitem{}
Reinecke, M., Hillebrandt, W., Niemeyer, J. C., Klein, R., and
Gr\"obl, A. 1998b, \aap, in press

\bibitem{}
Shy, S. S., Jang, R. H., and Ronney, P. D. 1996, Combust. Sci. Tech.,
113, 329 

\bibitem{}
Woosley, S. E. 1990, in Supernovae, ed. A. Petschek, (D. Reidel:
Dordrecht), 182 

\bibitem{}
Woosley, S. E. and Weaver, T. A. 1994, in Les Houches, Session LIV,
Supernovae, ed. S. Bludman, R. Mochkovitch, and J. Zinn-Justin
(Amsterdam: North-Holland), 63

\bibitem{}
Zeldovich, Ya. B. 1966, J. Appl. Mech. Tech. Phys., 7, 68

\bibitem{}
Zeldovich, Ya. B., Librovich, V. B., Makhviladze, G. M., and
Sivashinsky, G. L. 1970, Astronaut. Acta, 15, 313

\end{thebibliography}
\end{document}